\documentclass[fleqn,10pt]{article}

\usepackage[utf8]{inputenc}
\usepackage[T1]{fontenc}
\usepackage{mathtools}

\usepackage{authblk}

\usepackage{geometry}
\usepackage{setspace}
\usepackage{url}
\usepackage{amssymb}
\usepackage{booktabs}
\usepackage{nameref}
\usepackage{hyperref}
\usepackage{bbm}
\usepackage{placeins}

\usepackage{lineno}
\usepackage{cleveref}
\usepackage{xr-hyper}
\usepackage{pdfpages}
\usepackage{pgffor}
\def\supplementfilename{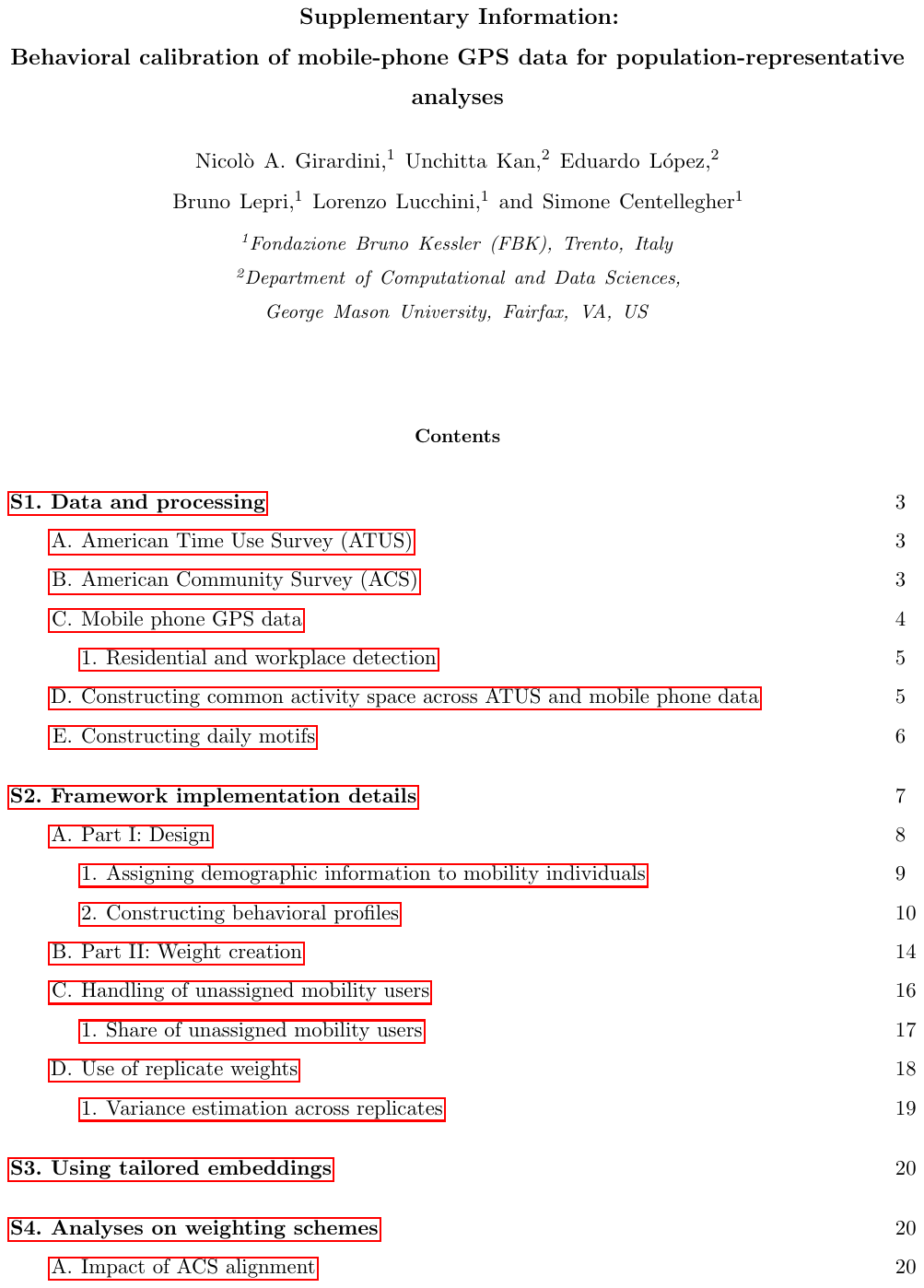}
\pdfximage{\supplementfilename}
\def\numbersupplementpages{\\the\pdflastximagepages}

\title{Behavioral calibration of mobile-phone GPS data for population-representative analyses}

\author[1,+]{Nicolò A. Girardini}
\author[2,+]{Unchitta Kan}
\author[2]{Eduardo L\'opez}
\author[1]{Bruno Lepri}
\author[1,$\dagger$,*]{Lorenzo Lucchini}
\author[1,$\dagger$]{Simone Centellegher}

\affil[1]{Fondazione Bruno Kessler (FBK), Trento, Italy}
\affil[2]{Department of Computational and Data Sciences, George Mason University, Fairfax, VA, US}

\affil[*]{llucchini@fbk.eu}
\affil[+]{These authors contributed equally to this work.}
\affil[$\dagger$]{Joint senior authors.}

\date{} % removes date

\begin{document}

\maketitle

\thispagestyle{empty}

\begin{abstract}
Mobile phone mobility data have transformed the study of human behavior, but demographic and behavioral biases can compromise their representativeness and distort population-level inference. Existing calibration approaches primarily address demographic and geographic representativeness, leaving behavioral discrepancies largely uncorrected. Here we introduce the Behavioral Population (BePop) framework, which jointly calibrates mobility data to representative demographic and behavioral distributions using census data and time-use surveys. BePop embeds mobility sequences into behavioral profiles and estimates person-level weights that align both population composition and daily activity patterns. Across three U.S. metropolitan areas, the framework consistently improves agreement between GPS-derived mobility and representative behavioral distributions, including time allocation, activity transitions, and mobility motifs. Calibration also substantially alters downstream mobility indicators, demonstrating that behavioral biases can propagate into commonly used mobility measures. Our results establish behavioral representativeness as a critical complement to demographic calibration and provide a general framework for population-representative mobility inference.
\end{abstract}


\section*{Introduction}
Mobile phone–derived location data have emerged as a powerful resource for understanding human behavior and its role in complex socio-dynamical systems. These intrinsically longitudinal and passively collected records provide high-resolution data streams that enable near-real-time monitoring and adaptive behavioral modeling, offering a dynamical perspective often unattainable through traditional surveys. Mobile phone data have already supported the study of fine-grained behavioral patterns across diverse domains, including human mobility and activity dynamics \cite{gonzalez2008understanding,pappalardo2015returners,alessandretti2020scales,schlapfer2021universal,bontorin2025mixing}; socioeconomic behavior, inequality, segregation, and crime \cite{singh2015money,dong2017social,luca2023crime,centellegher2025job,lucchini2025socioeconomic,chetty2022socialb,yabe2025behaviour,xu2025using,wang2016crime,song2019crime,albors2025deep}; and epidemiology and public health \cite{lucchini2021living,di2026mobility,oliver2020mobile,kraemer2020effect,aleta2020modelling}.
However, the rapid uptake and operational deployment of these data have outpaced systematic evaluations of their limitations. Mobile phone and GPS-based mobility data, although rich and timely, are susceptible to methodological challenges and biases that can distort scientific inference and misinform policy if left unaddressed~\cite{barreras2024exciting,yabe2024enhancing,kostandova2025improving}. These limitations include non-probability sampling and demographic representativeness biases, heterogeneity in data generation and measurement, and errors introduced during preprocessing~\cite{de2026howde,barreras2024exciting}.

The most widely discussed limitation concerns demographic representativeness, as mobile phone users constitute a non-probability sample of the population. Older adults, low-income communities, and certain geographic areas may be systematically under-represented because of differences in smartphone ownership, usage intensity, and data-sharing practices~\cite{coston2021leveraging,kostandova2025improving,yabe2024enhancing}. Data quality~\cite{wang2025exploring} and data generation~\cite{schlosser2021biases,den2026tale} may also vary across socio-demographic groups, potentially shifting aggregate mobility measures away from population-representative measures.

Additional biases can arise during preprocessing. Observed GPS pings are typically processed to identify stop locations or stays, which may subsequently be classified as home, work, or points of interest. GPS drift, limitations in complementary point of interest data, and differences among preprocessing pipelines can produce assignments that deviate from real visit patterns~\cite{barreras2024exciting,de2026howde,yabe2024enhancing}, leading to inconsistent or unrepresentative behavioral measures. Recent work has therefore sought to validate mobility-processing pipelines and the detection of different event types~\cite{de2026howde,li2025trajectory,wang2025cross,kato2024evaluating,pappalardo2021evaluation,huang2019transport,dodge2016analysis}.

Calibration has become a common strategy for mitigating representativeness bias in research using mobile phone data. The dominant approach relies on spatial post-stratification, linking device counts within geographic units to official population totals through expansion factors, under the assumption that observed devices are representative of residents within each unit~\cite{athey2021estimating,yabe2023behavioral,lucchini2025socioeconomic,centellegher2025job}. Although this approach can partially correct spatial imbalances in aggregate counts, it may not fully address demographic or behavioral biases. When geographic units are under-represented, absent, or suppressed because of device-count thresholds, spatial post-stratification may fail to recover reliable population estimates and may perpetuate biases affecting under-represented groups. More importantly, geographic calibration does not guarantee that behavioral proxies derived from GPS data align with those obtained from independent sources such as travel or time-use surveys. This misalignment remains potentially problematic even in analyses of relative change, particularly when the objective is to estimate population-representative behavioral quantities.

Here, we extend traditional survey-calibration methods~\cite{lohr2023sampling, sarndal2003model, deville1992calibration} to mobility data by introducing the Behavioral Population (BePop) framework, which jointly accounts for socio-demographic and behavioral dimensions. Moving beyond purely spatial or demographic post-stratification, BePop calibrates GPS-based mobility sequences using both population margins, derived from census data, and behavioral benchmarks, obtained from nationally representative time-use surveys widely used in behavioral research~\cite{robinson1999time,girardini2023adaptation,mcleod2023origins,rinderknecht2025daily}. The framework estimates individual-level weights that adjust the socio-demographic composition of the GPS sample to match that of the target population while simultaneously calibrating its behavioral composition against representative population distributions. Behavioral calibration targets are derived from individual activity sequences, represented with characteristics observable in both mobility and time-use datasets within a shared feature space. These sequences are subsequently clustered to identify distinct behavioral profiles.

We evaluate the framework using mobility data from three large U.S. metropolitan areas, with the American Time Use Survey (ATUS) providing the behavioral reference. We show that the joint calibration improves the alignment of GPS-derived behavioral distributions with the corresponding ATUS benchmarks. We further assess the impact of BePop weighting on downstream mobility indicators and longitudinal analyses, including periods of substantial behavioral change such as the COVID-19 pandemic.

We highlight that the BePop framework is compatible with established methods for weighted estimation and uncertainty quantification, and its behavioral representation can be adapted to the outcomes and reference data relevant to a given application. To facilitate its application to other mobility datasets and reference populations, we release a data-agnostic, open-source python implementation of the method. 

Taken together, BePop provides a flexible framework for mitigating socio-demographic and behavioral biases in mobile phone–derived data, enabling more robust population-representative inference in cross-sectional and longitudinal behavioral studies.

\section*{Results}

\subsection*{The BePop calibration framework}\label{subsec:framework}

\begin{figure}[htbp]
\centering
\includegraphics[width=1\linewidth]{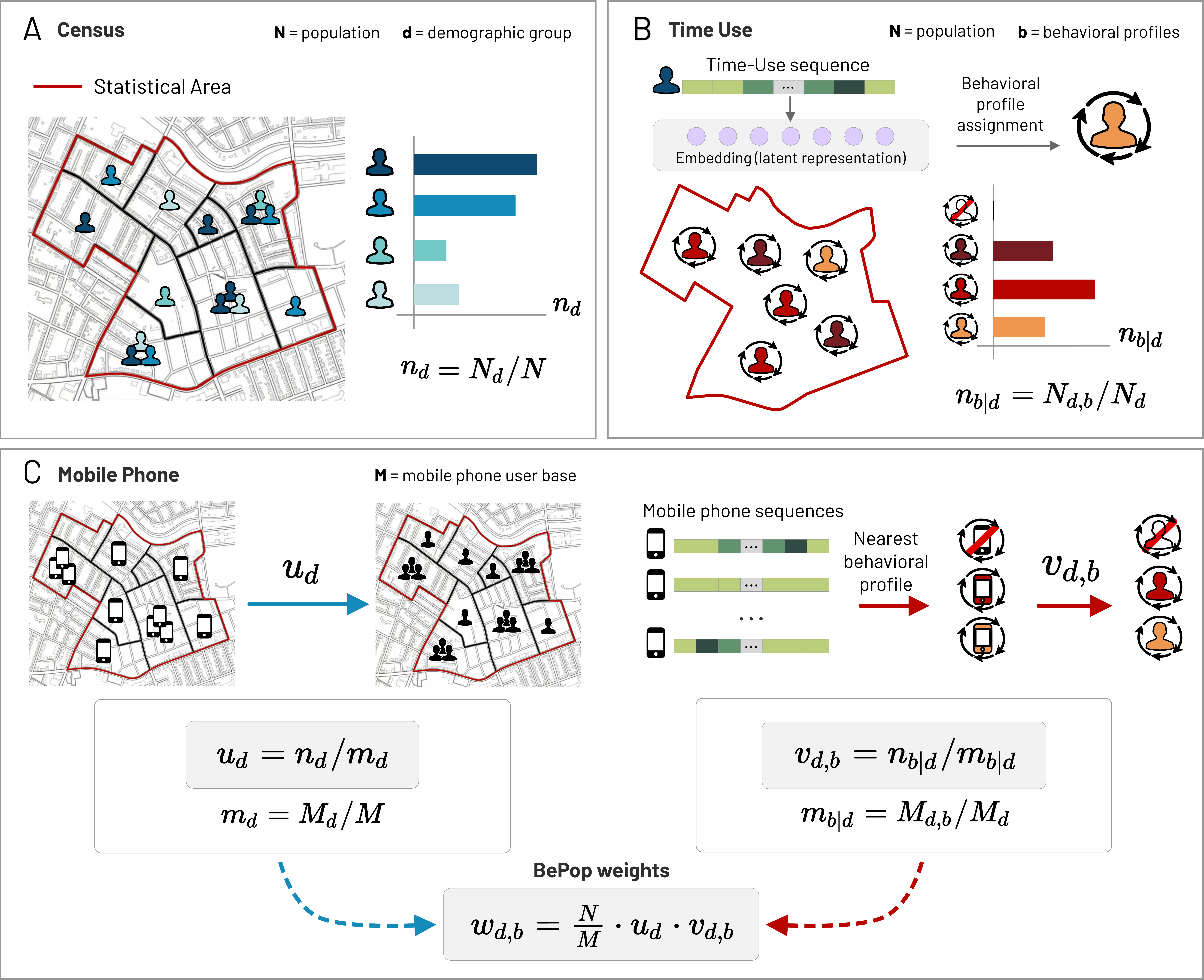}
\caption{\textbf{Overview of the BePop calibration framework.} The framework combines representative demographic and behavioral reference data to construct behavior-aware population weights for mobile phone users. \textbf{(A)} Census or community survey data define the population distribution across socio-demographic groups \textit{d} (e.g., age and income), yielding demographic targets $n_d = N_d/N$ within the statistical area of interest. \textbf{(B)} Time-use survey activity sequences are represented in a behavioral embedding space and assigned to a discrete set of behavioral profiles \textit{b}. Within each socio-demographic group, these profiles define conditional behavioral targets $n_{b|d}= N_{d,b}/N_d$. \textbf{(C)} Mobile phone users are first assigned to socio-demographic groups, allowing the computation of demographic adjustment factors $u_d = n_d/m_d$, where $m_d = M_d/M$ denotes the corresponding sample share. Daily mobility sequences are then matched to the nearest behavioral profile to estimate the sample behavioral composition $m_{b|d} = M_{d,b}/M_d$, from which behavioral adjustment factors $v_{d,b} = n_{b|d}/m_{b|d}$ are derived. The final BePop weight for users in demographic group \textit{d} and behavioral profile \textit{b} is given by $w_{d,b} = (N/M) \cdot u_d \cdot v_{d,b}$, simultaneously correcting demographic representation and behavioral composition while scaling the sample to the target population.
}
\label{fig:step_by_step}
\end{figure}

The Behavioral Population (BePop) weighting scheme is a calibration framework for mobility data that extends standard survey calibration approaches to jointly account for socio-demographic and behavioral biases in large-scale mobility samples. The core aim of calibration methods is to align the distribution of selected attributes observed in a sample with corresponding distributions in a reference or target population. 
The key idea behind BePop is to exploit two complementary reference data sources: demographic information from census data and behavioral information from representative time-use surveys. As illustrated in Fig.~\ref{fig:step_by_step}, the BePop framework jointly aligns mobile phone users with demographic population distributions and behavioral distributions conditioned on demographic groups, ultimately producing individual calibration weights. These weights can be interpreted as the number of individuals in the target population represented by each observed mobile phone user.

The framework consists of two main components: (i) the identification and construction of demographic and behavioral attributes that can be consistently defined across both the mobile phone and the reference datasets, and (ii) the estimation of calibration weights that align the resulting attribute distributions with the selected demographic and behavioral targets.

 \paragraph{BePop in practice: an application to US metropolitan areas.}
In this work, as a real-world application, we apply the BePop framework to a large longitudinal dataset of privacy-enhanced GPS mobility traces collected in the Phoenix ($\approx140$k users), Boston ($\approx96$k users), and New York ($\approx527$k users) metropolitan areas between January and February 2020. As reference demographic and behavioral information, we use data from the American Community Survey (ACS)~\cite{acs} and the American Time Use Survey (ATUS)~\cite{bls2026atus}, respectively, for the same metropolitan areas. Specifically, we used a sample size for the ATUS data of $\approx 2$k in Phoenix, $\approx 2.6$k in Boston, and $\approx 9$k in New York (see Methods and SI Sec.~\ref{sec:si_dataset_processing} for more details on the datasets).

To ensure comparability between mobile phone and ATUS data, we represent each day as a temporal sequence of 48 half-hour bins, with each bin representing the main activity performed during that interval. ATUS sequences are directly constructed from respondents' reported daily locations. 

\FloatBarrier

In contrast, mobility sequences are obtained by identifying stop locations, assigning them to activity categories through nearby points of interest (POIs), and mapping them onto the same temporal vector representation. A shared activity classification is adopted across the two datasets to ensure temporal and semantic alignment (see SI Table~\ref{tab:si_loc_mapping}).
For the calibration procedure, we randomly select one day of activity for each mobile phone user from the two months preceding the COVID-19 pandemic (January 3--February 29, 2020), yielding a final mobile phone sample of approximately 760{,}000 user-day observations across the three metropolitan areas. Additional details on data preprocessing, sequence construction, and the activity mapping are provided in the Methods and SI Sec.~\ref{sec:si_dataset_processing}.

\paragraph{Socio-demographic targets.} 
As shown in Fig.~\ref{fig:step_by_step}A, the census-based calibration targets are derived from ACS as the proportion of the population within each socio-demographic stratum $d$, $n_d=N_d/N$. Here $N_d$ represents the population of stratum $d$ and $N$ is the resident population within the statistical area of interest (e.g, a metropolitan area).

\paragraph{Behavioral targets.} 
Behavioral calibration targets are derived from ATUS. Daily activity sequences are first represented in a behavioral embedding space and then clustered into a discrete set of behavioral profiles $b$ (see Methods for details on the behavioral profile construction). 
Within each socio-demographic stratum $d$, the resulting distribution of behavioral profiles defines the conditional behavioral target proportions, $n_{b|d}=N_{d,b}/N_d$, where $N_{d,b}$ represents the number of individuals in demographic stratum $d$ assigned to behavioral profile $b$, and $N_d$ represents the population of stratum $d$ residing in the statistical area under study (Fig.~\ref{fig:step_by_step}B). 
Note that, by construction, behavioral targets are conditioned on the different population strata to preserve possible group-specific behavioral heterogeneities.

\paragraph{BePop calibration of mobile phone data.}
Since mobile phone data typically contain passively collected spatiotemporal information but no directly observed socio-demographic or behavioral attributes, these attributes were assigned to each user before calibration. In this work, socio-demographic attributes are inferred using established procedures based on users' inferred residential locations and the corresponding census information \cite{lucchini2021living,centellegher2025job,lucchini2025socioeconomic,de2026howde}, while behavioral attributes are obtained by matching mobile phone sequences to the reference behavioral profiles derived from ATUS (see Methods and SI Sec.~\ref{subsec:si_attributes}). The matching procedure allows mobile phone sequences that are inconsistent with any reference behavioral profile to remain \emph{unassigned}. Analysis indicates that these unassigned sequences primarily reflect low-quality mobility traces or mislabeled home, work, or POI locations, effectively making this step a quality filter (see SI Sec.~\ref{subsec:si_handling_unassigned} for more details).

Once each mobile phone user has been associated with a socio-demographic stratum $d$ and a behavioral profile $b$, the calibration estimates two sets of adjustment factors (Fig.~\ref{fig:step_by_step}C). Demographic adjustment factors align the mobility sample with the ACS population targets and are defined as $u_d=n_d/m_d$, where $n_d$ and $m_d$ denote the demographic proportions in the target population and the mobile phone sample, respectively. Behavioral adjustment factors are estimated independently within each socio-demographic stratum by comparing the behavioral composition of the mobility sample with the corresponding ATUS reference distribution, yielding $v_{d,b}=n_{b|d}/m_{b|d}$, where $n_{b|d}$ and $m_{b|d}$ represent the proportions of behavioral profiles in the ATUS reference data and the mobile phone sample, respectively.

The BePop weight assigned to each mobile phone user is obtained by combining the adjustment factors: $w_{d,b}=\frac{N}{M}\,u_d\,v_{d,b}$,
where the scaling factor $N/M$ expands the calibrated sample to the size of the target population (see Fig.~\ref{fig:step_by_step}C). 
The resulting weighted mobility sample simultaneously reproduces both the ACS socio-demographic distribution and the ATUS behavioral profile distribution conditional on each socio-demographic stratum. 
Further details are provided in the Methods section.

\subsection*{Evaluating the effectiveness of BePop calibration}

As the goal of the framework is to align mobility samples with behavioral targets and produce more representative estimates of population-level daily behavior, we evaluate the extent to which BePop calibration reduces discrepancies between the mobility sample and the reference datasets.
In this section, we report post-calibration aggregated results for the three metropolitan areas, alongside pre-calibration mobile phone and reference ATUS distributions.

\begin{figure}[ht]
\centering
\includegraphics[width=1\linewidth]{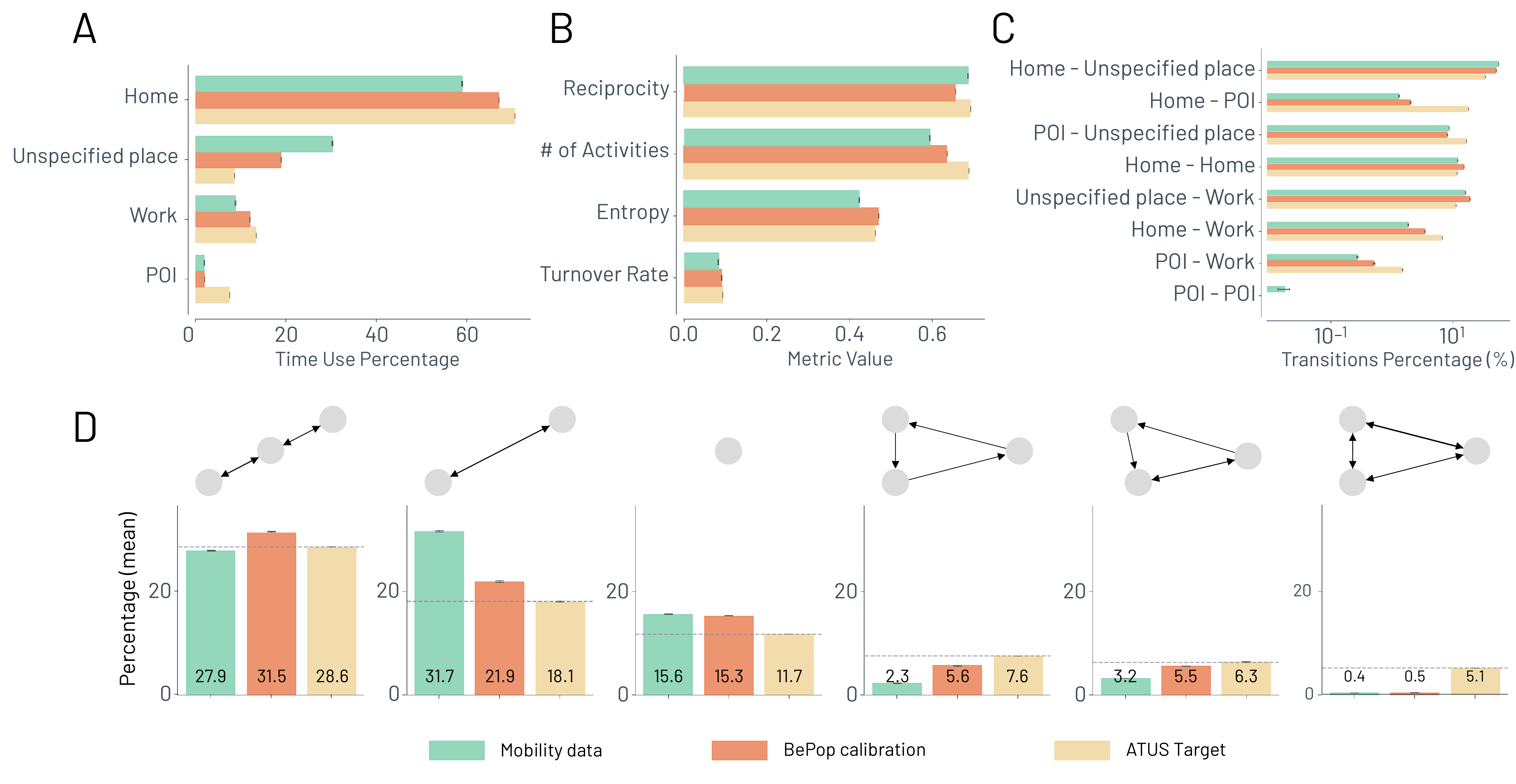}
\caption{\textbf{Alignment of calibrated mobility data with ATUS behavioral targets.} The panels report the alignment on three behavioral dimensions mirroring the embeddings construction: \textbf{(A)} Time-use, \textbf{(B)} Structural Metrics of activity sequences, \textbf{(C)} Transition Patterns. We also include \textbf{(D)} motifs constructed on transitions between activities. Overall, the \emph{BePop calibration} framework systematically increases the alignment of mobility data with the target data.}
\label{fig:metrics_results}
\end{figure}

First, we evaluate \emph{time allocation}, comparing the proportions of time spent across the main activity categories, i.e., \emph{Home}, \emph{Work}, \emph{POIs}, and \emph{Unspecified} (see Fig.~\ref{fig:metrics_results}A). Second, we assess \emph{structural metrics of activity sequences} by comparing sequence-level behavioral metrics, including reciprocity, the number of unique activities, turnover rate, and normalized entropy (see Fig.~\ref{fig:metrics_results}B). Third, we evaluate \emph{activity transition patterns} by comparing the proportions of transitions between activity categories within a sequence (see Fig.~\ref{fig:metrics_results}C). These three sets of metrics constitute the behavioral dimensions used to construct the behavioral embeddings for calibration (see Methods section).
Finally, we compare the frequency of \emph{daily behavioral motifs}, which summarize the structure of individuals' daily activity sequences (see Fig.~\ref{fig:metrics_results}D). Motifs are defined as directed networks describing the complete set of transitions across an individual's daily activities.
We refer to the Methods section for a detailed definition of all evaluation metrics and to the SI Sec.~\ref{subsec:si_motif_construction} for more details on motif construction.

All metrics are computed over 50 bootstrap replicates for each of the three metropolitan areas considered in the analysis (Phoenix, New York, and Boston), where each replicate is obtained by resampling mobile user-days with replacement. Results are reported as the mean value across bootstrap replicates together with 95\% confidence intervals. For the calibrated samples, estimates are computed using the BePop calibration weights $w_{d,b}$. These comparisons allow us to quantify the extent to which calibration improves the agreement between mobility-derived and reference population-level (ATUS reference) behavioral distributions.

In Fig.~\ref{fig:metrics_results}A, the uncalibrated mobile phone sample shows noticeable discrepancies with respect to the ATUS reference time-use distributions.
The calibration is highly effective in reducing these discrepancies, bringing the calibrated distribution of time-use closer to the ATUS baseline.
The BePop calibration shows a similar impact on sequence-level structural metrics. As shown in Fig.~\ref{fig:metrics_results}B, reciprocity and turnover rates are already well aligned in the uncalibrated sample and therefore exhibit consistent results also after calibration. In contrast, significant improvements are produced after calibration in the distributions of the number of activities and sequence entropy. 
Fig.~\ref{fig:metrics_results}C compares the distributions of the most common undirected transitions between location types. Also for this set of measures, we find the BePop calibration improves the alignment towards the ATUS distributions for nearly all transition categories.

In addition, we also evaluate the effect of calibration on metrics that are not explicitly included in the embedding space. Fig.~\ref{fig:metrics_results}D evaluates the prevalence within the population of different behavioral motifs, each representing daily activity sequences. We report the distributions of the proportion of the top six motifs, ranked by prevalence in the ATUS reference data. While these motifs are not explicitly included in the behavioral embedding used for calibration, we nevertheless find that, overall, the calibration improves their agreement with the reference distribution. 

Across cities and behavioral dimensions, comparing the original and BePop-weighted mobility samples against ATUS showed that BePop reduced the mean Jensen–Shannon (JS) distance by $9.71\% \pm 0.74$ ($14.80\% \pm 0.09$ for time use, $13.69\% \pm 0.25$ for structural metrics and $0.65\% \pm 2.21$ for transitions). Similar improvements were observed for motif distributions, despite these not being represented in the embedding space, with discrepancies reduced by $17.00\% \pm 0.16$ (see SI Sec.~\ref{sec:si_support_tests_beh} for more details). Furthermore, comparing the JS distance of the mobility sample weighted with BePop to the sample weighted with a calibration solely based on socio-demographic attributes, we see a relative improvement in the alignment with ATUS of $10.44\% \pm 0.70$, and of $17.10\% \pm 0.15$ when considering the motif distributions. Further details are provided in SI Sec.~\ref{subsec:si_demo_weights} and Sec.~\ref{sec:si_support_tests_beh}).

Results across different metropolitan areas show consistent effects of calibration on behavioral alignment, with local differences based on the metric of interest (see SI Sec.~\ref{sec:si_city_wise_calibration}). This suggests that careful design of the embedding space is required when the aim is to reproduce specific behavioral quantities for specific geographical areas (see SI Sec.~\ref{sec:si_embeddings} for detailed ``tailored embeddings'' results).

\subsection*{BePop calibration on longitudinal mobility data}

BePop weights can be computed for a changing observed population, allowing the calibration to adapt to temporal changes in the composition of the mobile phone sample. Longitudinal application, however, requires each user to be consistently associated with a socio-demographic stratum $d$ and a behavioral profile $b$ over time (see Methods Sec.~``BePop longitudinal calibration''). Here, we adopt the simplest approach, assuming that i) the socio-demographic stratum $d$ is uniquely determined during preprocessing, and ii) the behavioral profile $b$ can be consistently attributed based on a single randomly selected day in the reference period. This approach assumes that the mobile phone user reference-day profile captures a sufficiently persistent component of individual behavior to remain informative over time. We evaluate this assumption at both the population and individual levels and then examine how BePop calibration affects longitudinal mobility estimates. We assess the stability of this assumption by bootstrapping the user's reference-day and socio-demographic stratum assignments.

\paragraph{Population-level stability.}
We first assess whether fixed reference-profile assignments allow the calibrated population to remain responsive to changes in aggregate behavior. If the reference profiles capture persistent behavioral components, the corresponding weights should reproduce the target pre-pandemic ATUS behavioral composition while still allowing the weighted population to depart from this baseline when population-level behavior changes.

\begin{figure}[t!]
\centering
\includegraphics[width=0.95\linewidth]{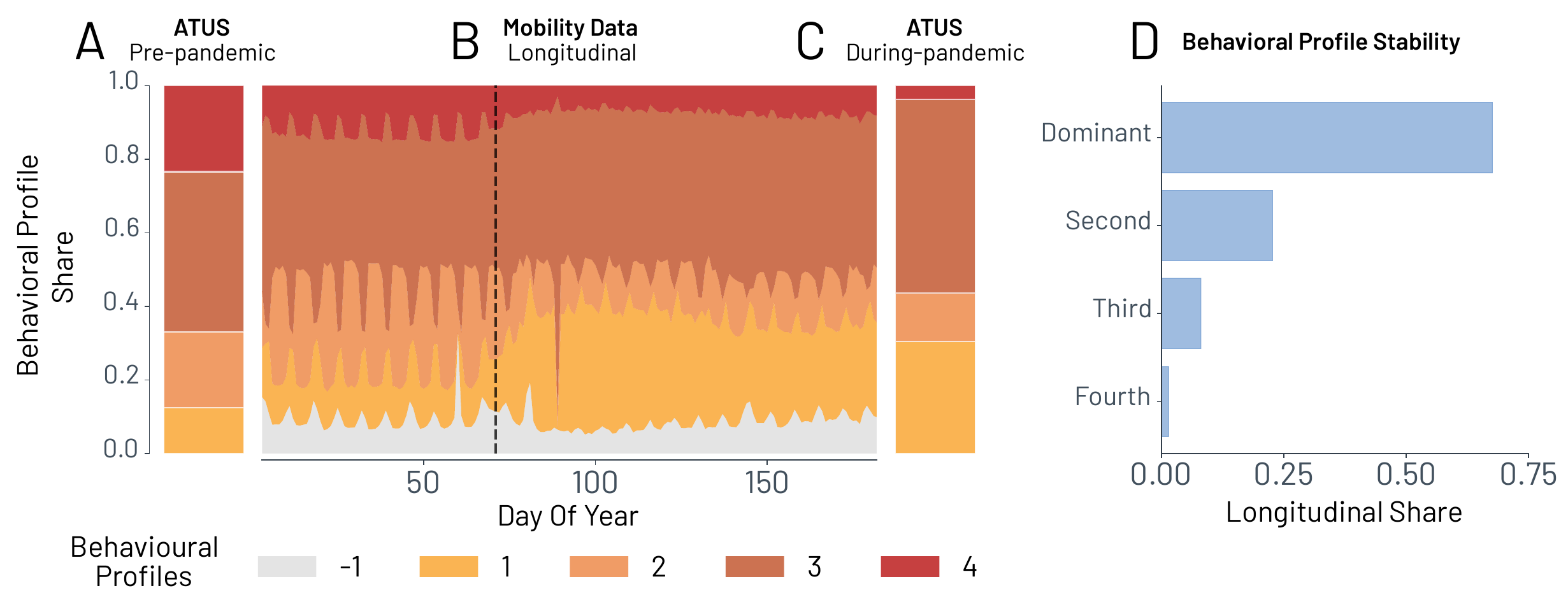}
\caption{\textbf{Longitudinal stability of BePop behavioral profiles (Phoenix).}
\textbf{(A)} Behavioral-profile distribution in the pre-pandemic ATUS sample, used as the reference for calibration.
\textbf{(B)} Temporal evolution of behavioral profile distributions in the BePop-weighted mobile phone sample. Each user is assigned a fixed reference behavioral profile inferred from a randomly selected pre-pandemic day, while calibration weights are recomputed daily using the active mobile phone sample. For validation, behavioral profiles are independently re-inferred from each user's observed mobility sequence on every day and aggregated using the corresponding day-specific weights. The dashed line marks the WHO pandemic declaration on March~11,~2020. For visual purposes, only one bootstrap realization of the users' reference-day and socio-demographic is shown.
\textbf{(C)} Behavioral profile distribution in the ATUS sample collected between May and August 2020, shown for qualitative comparison with the post-pandemic weighted mobility distribution.
\textbf{(D)} Individual-level stability of behavioral profile assignments during the pre-pandemic period. For each user, behavioral profiles inferred across observed days are ranked according to their assignment frequency. Bars report the mean fraction of observed days assigned to the dominant, second, third, and fourth most frequent behavioral profiles across users.
}
\label{fig:stability}
\end{figure}

For this analysis, each user's behavioral profile is independently re-inferred from the mobility sequence observed on each day. These daily profiles are then aggregated using weights computed from the user's fixed reference profile. We present results for the Phoenix metropolitan area in Fig.~\ref{fig:stability}. Similar results are obtained for the Boston and New York metropolitan areas (see SI Sec.~\ref{sec:support_longitudinal_analysis}).

The pre-pandemic ATUS behavioral profile reference marginals (Fig.~\ref{fig:stability}A), computed before the WHO pandemic declaration (March 11, 2020), serve as the benchmark for comparison with the BePop-weighted distribution of behavioral profiles inferred daily from mobility data (Fig.~\ref{fig:stability}B). The weighted distribution shows reasonable agreement with the ATUS reference. The observed weekday-weekend fluctuations are expected because the ATUS reference combines weekday and weekend diaries. Relative to the unweighted mobility data, BePop reduces the average Jensen-Shannon distance from the ATUS reference by $9.70\% \pm 0.61$ across bootstraps.

Following the pandemic declaration, the weighted distribution of behavioral profiles shifts away from its pre-pandemic composition. The direction of this shift is consistent with the changes observed in the distribution of behavioral profiles drawn from an independent ATUS sample (51 users for Phoenix) collected between May and August 2020 (Fig.~\ref{fig:stability}C). Note that all sequences are always mapped onto the pre-pandemic ATUS behavioral profile reference (Fig.~\ref{fig:stability}A).

Specifically, focusing on the pandemic data for both mobile phone and ATUS (both after May 2020 for consistency), the Jensen-Shannon distance computed between the weighted distribution and the ATUS distribution is reduced by $0.64\% \pm 3.86$ relative to the Jensen-Shannon distance computed between the unweighted distribution and the ATUS distribution (Fig.~\ref{fig:stability}B and C).
Because the pandemic ATUS sample is substantially smaller than the pre-pandemic sample, we also apply the bootstrap procedure to the ATUS sample when computing these distances. The resulting large standard errors prevent us from determining whether, and to what extent, the improvement associated with BePop persists during the pandemic period. Accordingly, these results should be interpreted as evidence of external consistency rather than as exact validation. Further details are provided in SI Sec.~\ref{sec:support_longitudinal_analysis}.

\paragraph{Individual-level stability.}
We next assess whether a mobility user behavioral profile inferred from a single reference day captures a sufficiently persistent component of individual behavior. For each user, behavioral profiles inferred across multiple observed days are ranked according to their assignment frequencies. The dominant profile accounts, on average, for more than $70\%$ of daily assignments (Fig.~\ref{fig:stability}D), indicating that the reference-day assignment captures a persistent component of individual behavior rather than only transient day-to-day variation.

Users whose reference-day sequence cannot be assigned to a behavioral profile receive zero weight and are therefore excluded from the longitudinal estimates. An included user can nevertheless be classified as unassigned on another day after the daily behavioral profile is re-inferred. Such unassigned classifications account for less than $10\%$ of evaluated user-days in both the pre-pandemic and pandemic periods (Fig.~\ref{fig:stability}). This indicates that the reference-profile assignments remain broadly compatible with users' subsequent daily behavior. For further details refer to SI Sec.~\ref{sec:support_longitudinal_analysis}).

\paragraph{Longitudinal mobility outcomes.}
The population and individual-level stability results indicate that fixed reference-profile assignments provide a sufficiently stable basis for longitudinal calibration while allowing the weighted population to remain responsive to systemic temporal changes in behavior. Having established the temporal consistency of the weighting strategy, we next examine how BePop calibration affects longitudinal mobility indicators.

We analyze daily mobility from January through June 2020, spanning both the pre-pandemic period and the initial COVID-19 lockdowns. Our objective is not to estimate the causal effect of the pandemic, but rather to quantify how correcting socio-demographic and behavioral composition biases influences longitudinal mobility estimates. We consider two commonly used mobility indicators: the radius of gyration~\cite{gonzalez2008understanding} and the frequency of co-location events~\cite{lucchini2021living}. Each metric is computed daily and averaged across users using the day-specific BePop weights, $w_{d,b}(t)$. We report weighted means and 95\% confidence intervals separately for Phoenix, Boston, and New York, together with their uncalibrated counterparts.

\begin{figure}[ht!]
\centering
\includegraphics[width=0.95\linewidth]{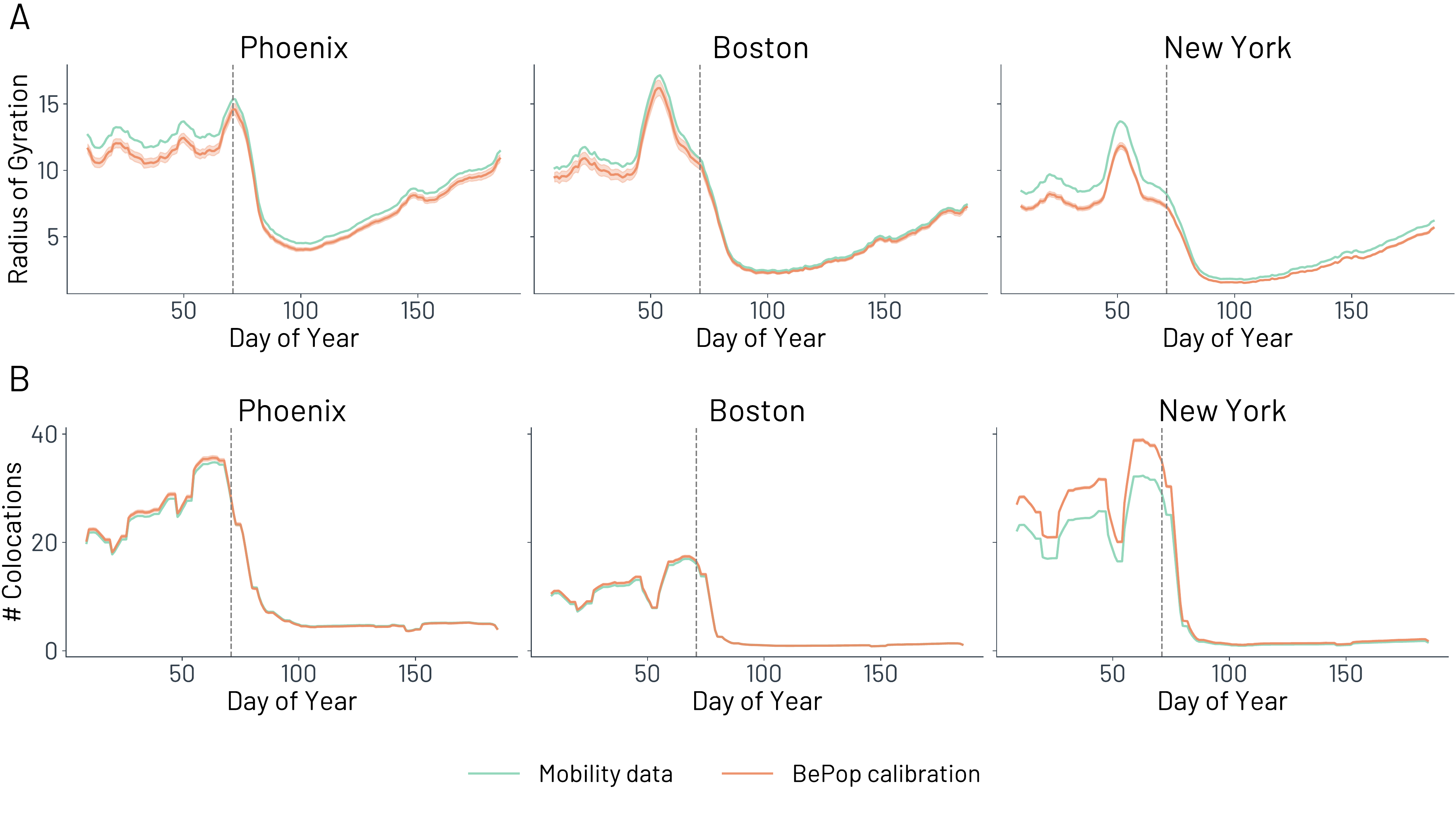}
\caption{\textbf{Effect of BePop weighting on longitudinal mobility metrics.}
\textbf{(A)} Daily mean radius of gyration in Phoenix, Boston, and New York.
\textbf{(B)} Daily mean number of co-location events at business locations. Curves compare estimates obtained from the uncalibrated mobile phone sample with those obtained using BePop weights. Both time series are shown as 7-day rolling averages, and intervals indicate 95\% confidence intervals obtained by bootstrap resampling. The vertical dashed line marks the WHO pandemic declaration on March 11, 2020.}
\label{fig:covid_results}
\end{figure}

Fig.~\ref{fig:covid_results}A shows the temporal evolution of the radius of gyration. Mobility declines sharply in all three metropolitan areas following the WHO pandemic declaration. BePop-weighted estimates remain systematically lower than the corresponding uncalibrated estimates. During the pre-lockdown period, the reduction is $9.03\% \pm 2.20$ in Phoenix, $6.74\% \pm 1.96$ in Boston, and $14.29\% \pm 1.80$ in New York ($10.02\%$ across cities). During the lockdown period, the corresponding reductions are $7.78\% \pm 2.28$ in Phoenix, $5.18\% \pm 1.98$ in Boston, and $11.86\% \pm 1.79$ in New York ($8.27\%$ across cities). Although both approaches capture a similar temporal contraction in mobility, calibration substantially changes the estimated absolute level of individual mobility.

A contrasting pattern emerges for business co-location events, which provide a proxy for shared physical presence and potential social interaction~\cite{lucchini2021living}. Fig.~\ref{fig:covid_results}B shows that BePop calibration produces higher estimates of business co-location events during the pre-lockdown period. During this period, calibration increases the estimated number of business co-location events by $2.32\% \pm 0.86$ in Phoenix, $3.78\% \pm 1.09$ in Boston, and $23.99\% \pm 3.17$ in New York ($10.03\%$ across cities). Following the pandemic declaration, both calibrated and uncalibrated estimates decline rapidly and converge toward similarly low absolute number of co-location events. During the lockdown period, calibration changes the estimated number of business co-location events by $-1.69\% \pm 1.19$ in Phoenix, $0.20\% \pm 1.49$ in Boston, and $17.95\% \pm 2.34$ in New York ($5.49\%$ across cities).

Because the calibrated estimates begin from a higher pre-pandemic baseline while converging to similar levels during lockdown, they imply a larger relative reduction in business-related interactions. More generally, these results demonstrate that calibration can affect both the estimated level of a longitudinal mobility indicator and the magnitude of its temporal change. Depending on the outcome considered, analyses based on uncalibrated mobile phone data may therefore either underestimate or overestimate changes in population mobility.

%%%%%%%%%%%%%%%%%%%%%
\section*{Discussion}

Existing calibration approaches for mobile phone-derived data have focused mainly on improving demographic and geographic representativeness through techniques such as spatial post-stratification \cite{centellegher2025job,lucchini2025socioeconomic,yabe2023behavioral, athey2021estimating}. Although effective in correcting spatial and demographic imbalances, these methods do not directly address potential behavioral imbalances caused by the heterogeneous usage of recording devices. Even after aligning the demographic composition of the mobile phone sample with census populations, systematic discrepancies in activity patterns might remain (see SI Sec.~\ref{subsec:si_demo_weights}). This highlights the need to explicitly calibrate behavioral characteristics when the objective is to obtain population-representative estimates of daily human behavior.

In this work, we introduced the Behavioral Population (BePop) framework, a calibration methodology that jointly aligns mobile phone data with population-representative socio-demographic and behavioral benchmarks. By combining census-derived demographic margins with behavioral profiles extracted from nationally representative time-use surveys, BePop extends traditional survey calibration to account for dimensions of human behavior that are typically overlooked in mobility data.

Across three metropolitan areas, we showed that joint calibration consistently improves the agreement between GPS-derived mobility sequences and representative behavioral distributions, including time allocation, activity-sequence structure, transition patterns, and behavioral motifs. BePop calibration also affects downstream mobility indicators, such as the radius of gyration and workplace co-locations, both before and during the COVID-19 pandemic. These results suggest that analyses relying exclusively on raw mobility data may systematically affect estimates derived from such data sources.

An important feature of BePop is its flexibility. The framework design is consistent with established survey calibration methodologies~\cite{deville1992calibration, sarndal2003model, breidt2017model} and assigns a single weight to each mobile phone user-day. It can accommodate different mobility datasets, spatial and temporal resolutions, demographic classifications, and application-specific behavioral benchmarks. Although we demonstrate the approach using a specific GPS mobility dataset together with ACS and ATUS references, the methodology is largely data-agnostic and not tied to a particular dataset or country.  

Our analyses also highlight the importance of behavioral representations in calibration. The behavioral embedding determines which aspects of daily activity are preserved and therefore which behavioral dimensions are best aligned after calibration. As shown in SI Sec.~\ref{sec:si_embeddings}, no single embedding optimizes every behavioral measure; instead, application-specific embeddings consistently perform best for their targeted behavioral dimensions. Richer embeddings capturing multiple aspects of behavior may therefore offer a practical compromise between generality and calibration performance.

Several limitations remain. As with any calibration methodology, the quality of calibrated estimates depends on the representativeness and measurement quality of the reference data, while sparse or high-dimensional behavioral targets may reduce calibration stability. 
The quality of mobile phone data also remains critical, particularly for longitudinal applications. Although BePop weights are estimated at the user-day level, extending behavioral profiles across multiple days requires sufficient temporal consistency in the underlying observations. This limitation is intrinsic to mobile phone data and should be considered when interpreting results, especially during periods of abrupt behavioral change such as the COVID-19 pandemic. At the same time, persistent failures in behavioral profile assignment may provide a useful quality-control signal for identifying low-quality data.

Overall, our work argues for a broader perspective on representativeness in mobility research. While demographic representativeness has long been recognized as an important requirement for population inference, our results show that it does not necessarily imply a balanced recording of user activities and, in turn, behavioral representativeness. Behavioral biases may persist even in demographically calibrated datasets and can substantially influence downstream conclusions. By explicitly integrating representative behavioral benchmarks into the calibration process, the BePop framework provides a practical methodology for reducing these discrepancies while remaining compatible with established survey-weighting principles. This work represents a step towards considering behavioral representativeness as a complementary criterion for assessing data quality and to incorporate behavioral calibration as a standard component of more robust mobile phone data analyses.

\section*{Methods}
\label{sec:methods}

\subsection*{Data overview}

The Behavioral Population (BePop) calibration framework leverages population-level socio-demographic and time-use information to de-bias mobile phone location data. In our analysis, as a real-world application, we use three complementary data sources: 
\begin{itemize}
    \item U.S. Census American Community Survey (ACS) data provide age and income distributions. These are used both (i) to probabilistically assign mobile phone users to demographic groups, at the Census Block Group (CBG) level, and (ii) to compute calibration targets for the Phoenix, Boston, and New York Core-Based Statistical Areas (CBSAs).
    \item Behavioral reference patterns are derived from the American Time Use Survey (ATUS), a nationally representative 24-hour activity-diary survey with respondent weights and sociodemographic information using ATUS records from 2004–2019 for each of the three CBSAs (Phoenix, Boston, and New York). For the longitudinal validation, we additionally use ATUS diaries collected between May and August 2020 to characterize time-use after the WHO pandemic declaration. 
    \item A dataset of privacy-enhanced mobile phone GPS data in the period from January to June 2020, from users that opted-in to share the data anonymously. The data collection is compliant with both the General Data Protection Regulation (GDPR) and the California Consumer Privacy Act (CCPA). We include users from three large metropolitan areas, namely Phoenix ($\approx140$k users after preprocessing), Boston ($\approx96$k users), and New York ($\approx527$k users). Only mobility data between January and February 2020 are used to compute behavioral profiles. 
\end{itemize}

Further details are provided in SI~\ref{sec:si_dataset_processing}.

\subsection*{Activity Sequences}

From the GPS traces, we identify stops, infer home and work locations from recurring nighttime and weekday daytime activity, and classify other stops types by linking them to nearby OpenStreetMap points of interest (POIs) \cite{lucchini2021living,centellegher2025job}, excluding sensitive locations (see SI Sec.~\ref{subsec:si_activity_seq} for details on the data preprocessing procedures).

In order to align ATUS activity data with GPS-based activity data, we define a mapping between the types of location identified in ATUS and the types of attributed stops in the mobile phone data (see SI Table~\ref{tab:si_loc_mapping}). For this mapping, we assume the gaps between stop locations in the mobility data to be time spent in \emph{Unspecified places}. With a similar line of reasoning, ATUS activities classified under \emph{Travel} are also reassigned to the \emph{Unspecified place} category.

For both datasets, we construct daily temporal sequences of activities represented as vectors of 48 time bins, each corresponding to a 30-minute interval (see SI Sec.~\ref{subsec:si_activity_seq} for further details). 
In the following, we refer to ATUS--derived sequences as $s$ and $\sigma$ for mobile phone-derived sequences.

\subsection*{BePop Calibration Framework}
\label{sec:framework-design}

The BePop framework jointly calibrates the mobile phone sample across both socio-demographic and behavioral dimensions. This requires first defining a set of calibration attributes for which reliable population or survey-based targets are available, and then assigning the corresponding attributes to mobile phone users. These attributes are then used to compute behavior-aware socio-demographic weights.

\subsubsection*{Socio-demographic targets}

In this work, we consider age and income as demographic variables. Alternative demographic attributes may be adopted depending on the application or reference data available. We denote age by $a$ and income by $e$, and use the shorthand $d = (a,e)$ to refer to the specific demographic group.

Specifically, we use four age groups, $a \in \{ 18-24, 25-44, 45-66, 67+\}$, and four income groups, $e \in \{\text{low}, \text{medium-low}, \text{medium-high}, \text{high}\}$, based on the quartiles of the income distribution of each CBSA. The design of socio-demographic groups is based on the specific task of interest. In our illustrative case, we selected groups that could capture different life course stages as well as labor force participation stages (e.g., students or young professionals, middle adulthood and potential parents, aging adults, retirees).
Note, that the same socio-demographic categorization scheme must be consistent across all data sources used in the calibration procedure. In particular, ATUS respondents within each CBSA must be mappable to the same age and income groups used to construct the ACS calibration targets. The same applies for the mobile phone data (see SI Sec.~\ref{subsec:si_assign_sociodemo} for details on the association procedure of demographic variables to mobile phone data).

Based on the identified socio-demographic groups, we define the population within a specific stratum $d$ as $N_d$. We denote the overall population of the area of interest (e.g. a CBSA) as $N$. Socio-demographic targets are then defined as: $n_d = N_d/N$.

\subsubsection*{Behavioral targets} 
In addition to socio-demographic attributes, we define behavioral attributes that summarize individual daily activity patterns. Since activity sequences are complex objects, we first represent each sequence through a behavioral vector embedding. 
The behavioral embedding is designed to capture different aspects of daily behavior, including activity durations, transition patterns, and the overall structure of the activity sequence. Alternative embeddings schemes may be adopted depending on the behavioral dimensions of interest and on the available data. More details on the construction of the behavioral embeddings are provided below.
%in Sec.~\ref{sec:method-vector-embeddings}. 
Sensitivity analyses using different embedding schemes are reported in SI Sec.~\ref{sec:si_embeddings}.

Specifically, we use the ATUS activity sequences $s$ to construct a discrete set of behavioral profiles. Pairwise distances between embedded ATUS sequences, $\delta(s,s')$, are computed using cosine distance. Similar sequences are grouped using K-Medoids clustering \cite{kaufmann1987clustering, park2009simple}. In this work, we define $K=4$ behavioral profiles, $b \in {1,\dots, K}$, where each profile represents a recurring pattern of daily activity behavior. Details on the behavioral profile construction are reported in SI Sec.~\ref{subsec:profiles_construction}. 

Based on the identified behavioral profiles, we denote the population within socio-demographic stratum $d$ and behavioral profile $b$ as $N_{d,b}$. The behavioral target is therefore defined by the distribution of ATUS-derived behavioral profiles within each socio-demographic stratum as $n_{b|d}=N_{d,b}/N_d$.

Note that in this case, the same behavioral categorization scheme must be also applied consistently across the mobile phone data. This is done by assigning each mobility sequence $\sigma$ to one of the ATUS behavioral profiles via a matching process. Explicitly, $\delta(s,\sigma)$ is calculated for all pairs of ATUS and mobile phone sequences. Then, for each mobile phone sequence $\sigma$ we consider the set of $k=10$ nearest ATUS neighbors and assign $\sigma$ to a behavioral profile $b$ if at least half of the neighbors belong to that profile (otherwise, $\sigma$ is left unassigned), following a majority consensus rule. Additionally, we also leave $\sigma$ unassigned if its distance from the medoid of the closest behavioral profile exceeds the $99th$ percentile of within-behavioral-profile distances. This avoids assigning behavioral profiles to mobile phone sequences that are ambiguous or too dissimilar from the ATUS observations defining the reference behavioral profiles.
We consider unassigned mobile phone sequences as low-quality data and treat those as zero-weight calibration strata in computation of the BePop weights (more in SI Sec.~\ref{subsec:si_handling_unassigned}).

\subsubsection*{BePop weights construction}\label{sec:framework-weights}
Each mobile phone user is now associated with a socio-demographic stratum $d$ and with a behavioral profile $b$. Let us then denote $M$ as the total number of users in the mobile phone sample within the area of interest. Let $M_d$ be the number of mobile phone users assigned to socio-demographic stratum $d$. Similarly, let $M_{d,b}$ denote the number of mobile phone users in stratum $d$ assigned to behavioral profile $b$. The corresponding target population quantities are denoted by $N$, $N_d$, and $N_{d,b}$.
The first factor contributing to the BePop weights, $u_d$, relates the mobile phone sample to the ACS socio-demographic targets. We define the sample and population shares of stratum $d$, respectively, as $$m_d = \frac{M_d}{M} \qquad \text{ and }\qquad n_d = \frac{N_d}{N}$$
Then, the socio-demographic adjustment factor $u_d$ is defined as
$$u_d =\frac{n_d}{m_d}$$

The second factor contributing to the BePop weights relates the behavioral composition of the mobile phone sample with the ATUS reference data within each socio-demographic stratum. For each stratum $d$, we define the conditional share of mobile phone users and population assigned to behavioral profile $b$ respectively as
$$m_{b|d} = \frac{M_{d,b}}{M_d} \qquad \text{ and } \qquad n_{b|d} = \frac{N_{d,b}}{N_d}$$
Here, $n_{b|d}$ is estimated from ATUS also accounting for the survey weights provided by the Bureau of Labor Statistics. Then, the behavioral adjustment factor is defined as
$$v_{d,b} = \frac{n_{b|d}}{m_{b|d}}$$

Finally, we can define the behavior-aware population weights (BePop weights) for a mobile phone user assigned to socio-demographic stratum $d$ and behavioral profile $b$ as
$$w_{d,b} = \frac{N}{M} \cdot u_d \cdot v_{d,b}$$

Note that factor $N/M$ scales the weights to the population size of the area of interest, so that each weighted mobile phone user can be interpreted as representing a number of individuals in the target population while $u_d$ and $v_{d,b}$ align the mobile phone sample with both reference data sources (in our example, ACS and ATUS). Within each socio-demographic stratum, the behavioral adjustment ensures that the weighted distribution of behavioral profiles matches the ATUS distribution of the same stratum. At the same time, because the behavioral adjustment is applied conditionally within each stratum $d$, the socio-demographic alignment to the ACS targets is preserved. The resulting weighted sample therefore reproduces the ACS socio-demographic distribution while also matching the ATUS-derived behavioral profile distribution within each socio-demographic stratum. For further technical details see SI Sec.~\ref{subsec:framework-weights}.

\subsubsection*{BePop longitudinal calibration}
Longitudinal application of BePop separates the assignment of users to calibration groups from the daily computation of their calibration weights. Each user is assigned to a socio-demographic stratum $d$ and a reference behavioral profile $b$ using a randomly sampled mobile user-day from January or February 2020. These assignments remain fixed throughout the analysis, while the calibration factors are recomputed independently on each day $t$ using the subset of assigned users observed on that day. This yields day-specific weights
$$w_{d,b}(t)=\frac{N}{M(t)}\frac{n_d}{m_{d}(t)}\frac{n_{d,b}}{m_{b|d}(t)}$$
where $m_{b|d}(t)$, $m_d(t)$, and $M(t)$ denote the day-specific behavioral, demographic, and sample populations, respectively. This procedure accounts for temporal changes in the composition of the observed mobile phone sample while allowing users' daily mobility behavior to evolve over time.

To evaluate the longitudinal use of BePop, we restrict the analysis to users observed both before and after the WHO pandemic declaration on March 11, 2020. This restriction limits changes in sample composition caused by complete panel dropout, although it defines a persistent-user cohort and may therefore introduce survivorship bias. Additional details on the longitudinal sample-selection criteria and weight construction are provided in SI Sec.~\ref{sec:support_longitudinal_analysis}. Here, we use a single-day assignment strategy to maintain consistency with the framework; developing and evaluating alternative longitudinal profiling methods is beyond the scope of this work.

\subsubsection*{Building behavioral embeddings from activity sequences}\label{sec:method-vector-embeddings}

For each activity sequence $s$ in the ATUS dataset and each sequence $\sigma$ in the mobile phone dataset, we construct a behavioral embedding vector that summarizes multiple aspects of daily behavior. In general, as we discussed, the different dimensions of the embedding space can be modeled to capture specific behavioral aspects that might be of particular relevance for the downstream analyses. In this work the behavioral embedding space is composed of features belonging to three complementary macro-categories, each capturing a different behavioral dimension: (\emph{i}) structural properties of the sequence, (\emph{ii}) time-use allocation, and (\emph{iii}) mobility transitions. Together, these features provide a compact representation of an individual's daily behavioral profile. We now describe each of these categories using $s$ as an example.\\
\\
\textbf{Structural metrics.} Let $s=(s_1,\dots,s_L)$ denote an activity sequence of length $L=|s|$, where $s_i$ is the activity observed at position $i$. These features are aimed at characterizing the structural properties of sequence $s$:
\begin{itemize}
    \item \emph{Number of Activities:} Number of distinct activities observed in sequence $s$, normalized by the total number of possible activity types $|A|$, where ($|A|=4$): \emph{Home}, \emph{Work}, \emph{POIs (Points of Interest)}, and \emph{Unspecified}. This encodes how many of the possible activities are included in sequence $s$.
    \item \emph{Turnover rate:} For each ordered pair of activity types $(x,y)$, we define the transition count: \[C_{xy}(s)=\sum_{i=1}^{L-1} \mathbf{1}_{s_i=x,\ s_{i+1}=y}.\] Thus, $C_{xy}(s)$ is the number of times activity $x$ is followed by activity $y$ in sequence $s$. The turnover rate measures how frequently the activity type changes between consecutive elements of the sequence. It is defined as the proportion of consecutive pairs involving two different activities:\[T(s)=\frac{\sum_{x\neq y} C_{xy}(s)}{L-1}.\]
    \item \emph{Reciprocity:} Measures the extent to which between-activity transitions occur in both directions, accounting for their relative frequencies. It is computed only over transitions between distinct activities. For each unordered pair of activity types $(x,y)$ occurring in $s$, the activity-level reciprocity component is given by the number of transitions that can be matched in both directions: $\min{C_{xy}(s),C_{yx}(s)}$. We define sequence-level reciprocity as the average activity-level reciprocity component over all observed between-activity transitions: \[R(s)=\frac{\sum_{x<y} \min{(C_{xy}(s),C_{yx}(s))}}{\sum_{x\neq y} C_{xy}(s)/2}.\] The measure ranges from $0$ to $1$, being equal to $1$ when transitions between all observed different activity pairs are perfectly balanced in the two directions. If no between-activity transitions are observed, $R(s)$ is set to $0$.
    \item \emph{Normalized Entropy:} We use the normalized Shannon entropy to quantify the diversity of performed activities from the activity set $A$, within a sequence $s$: \[H(s)= -\frac{\sum_{a \in A} p(a_i)\log p(a_i)}{\log |A|},\]
    where $p(a_i)$ is the empirical probability of observing activity $a_i$.\\
\end{itemize}

\noindent
\textbf{Time-use allocation.} These features describe how individuals distribute their time across different types of activities throughout the day. We include in these feature category the fraction of the day spent in each of the main activity categories: \emph{Home}, \emph{Work}, \emph{POIs (Points of Interests)}, and \emph{Unspecified}.\\
\\
\textbf{Transition features.} These features adopt a network perspective by treating each daily sequence as a motif that captures the structure of transitions between activities. In this feature category, we include normalized transition frequencies between all pairs of activities, including $x=y$ for days which only present a single activity (e.g., always at \emph{Home}). A comprehensive ``transition vocabulary'' is constructed from the combinations of all possible locations. Each sequence is represented as a vector over this shared set of transitions. Transitions that are not observed in a sequence are set to zero.\\
\\
The resulting behavioral embedding provides a flexible and interpretable representation of daily behavior. While the feature set adopted here is designed to capture complementary behavioral dimensions, the framework naturally supports the inclusion of additional features, allowing the embedding to be tailored to different datasets and analytical objectives. SI Sec.~\ref{sec:si_embeddings} reports the effect on behavioral metrics of BePop weights computed using different set of features when building the embedding space.

\bibliographystyle{plain}
\bibliography{sample}

\section*{Acknowledgements}
The authors would like to thank Cuebiq that kindly provided us with the mobility dataset for this research through their Data for Good program. This work was partially supported by ELIAS (Grant agreement ID: 101120237).

\section*{Author contributions statement}
L.L., S.C., N.A.G., U.K. conceived the original idea and planned the experiments. S.C., L.L. pre-processed the mobility data. N.A.G., U.K., S.C., L.L. carried out the experiments and made the figures. N.A.G., U.K., S.C., L.L., E.L., B.L. contributed to the interpretation of the results. L.L., S.C., N.A.G., U.K., wrote the manuscript. N.A.G., U.K., S.C., L.L., E.L., and B.L. provided critical feedback, helped shape the manuscript, and substantively revised it.

\section*{Data and Code availability}
The data supporting the findings of this study are accessible through Cuebiq’s Data for Good initiative. For details on how to request access, including conditions and limitations, please visit: \url{https://www.cuebiq.com/about/data-for-good/}.
American Time Use Survey (ATUS) and American Community Survey (ACS) data are both available for free online through official government and academic portals. 
Open-source implementation of the BePop calibration framework is available on GitHub at \url{https://github.com/unchitta/mob-calibrate}.
Any additional information required to reproduce the results of this paper is available from the lead contact upon request.

\includepdf[pages=-]{\supplementfilename}
\end{document}